\documentclass{msi}
\usepackage{amsmath}
  \usepackage{paralist}
  \usepackage{graphics} %% add this and next lines if pictures should be in esp format
  \usepackage{epsfig} %For pictures: screened artwork should be set up with an 85 or 100 line screen
\usepackage{graphicx}  \usepackage{epstopdf}%This is to transfer .eps figure to .pdf figure; please compile your paper using PDFLeTex or PDFTeXify.
 \usepackage[colorlinks=true]{hyperref}
\hypersetup{urlcolor=blue, citecolor=red}

\def\currentvolume{X}
 \def\currentissue{X}
  \def\currentyear{200X}
   \def\currentmonth{XX}
    \def\ppages{X--XX}
     \def\DOI{10.3934/xx.xx.xx.xx}

\theoremstyle{definition}

\DeclareMathOperator{\argmin}{argmin}

\title[Machine Learning for Yield Curve Feature Extraction] %Use the shortened version of the full title
      {Machine Learning for Yield Curve Feature Extraction: Application to Illiquid Corporate Bonds}

\author[Greg Kirczenow, Masoud Hashemi, Ali Fathi and Matt Davison]{}


 \email{gkirczen@rbc.com}
 \email{masoud.hashemi@rbc.com}
  \email{ali.fathi@rbccm.com}
  \email{mdavison@uwo.ca}
\thanks{$^*$All contents and opinions expressed in this document are solely those of the authors and do not represent the
view of RBC Financial Group. }

\begin{document}
\maketitle

% Enter the first author's name and address:
%\centerline{\scshape Greg Kirczenow$^*$}
%\medskip
%{\footnotesize
% please put the address of the first author
 %\centerline{RBC, Enterprise Model Risk Management}
   %\centerline{Other lines}
   %\centerline{ 200 Bay Street, ON, CA}
%} % Do not forget to end the {\footnotesize by the sign }

\medskip

\centerline{\scshape Greg Kirczenow$^*$ , Masoud  Hashemi$^*$, Ali Fathi$^*$ and Matt Davison}
\medskip
%{\footnotesize
 % please put the address of the second  and third author
 %\centerline{ First line of the address of the second author}
  % \centerline{Other lines}
   %\centerline{Springfield, MO 65810, USA}
%}

\bigskip

% The name of the associate editor will be entered by an editorial staff
% "Communicated by the associate editor name" is not needed for special issue.
 %\centerline{(Communicated by the associate editor name)}

%The abstract of your paper
\begin{abstract}
This paper studies an application of machine learning in extracting features from the historical market implied corporate bond yields.
We consider an example of a hypothetical illiquid fixed income market. After choosing a surrogate liquid market, we apply the Denoising Autoencoder (DAE) algorithm to learn the features of the missing yield parameters from the historical data of the instruments traded in the chosen liquid market. The DAE algorithm is then challenged by two "point-in-time" inpainting algorithms taken from the image processing and computer vision domain. It is observed that, when tested on unobserved rate surfaces, the DAE algorithm exhibits superior performance thanks to the features it has learned from the historical shapes of yield curves.
\end{abstract}

%The title of your section 1
\section{Introduction}

\begin{figure}[htp]
\begin{center}
  % replace aims_logo.eps by your figure file name
  \includegraphics[width=4in]{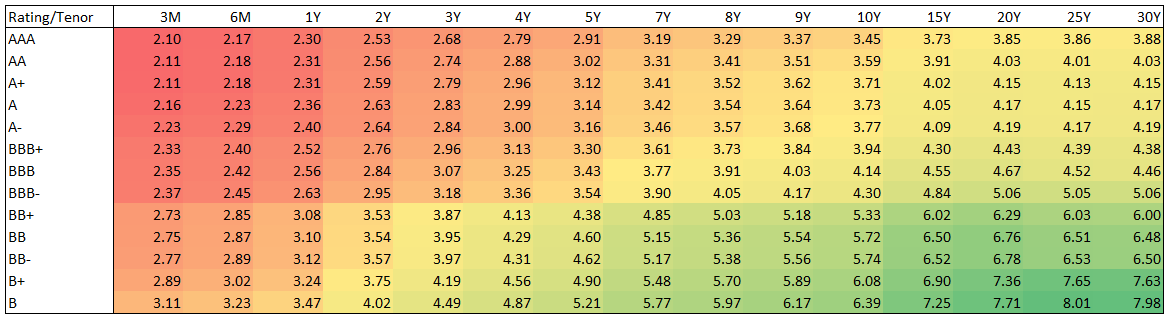}\\
  \includegraphics[width=4in]{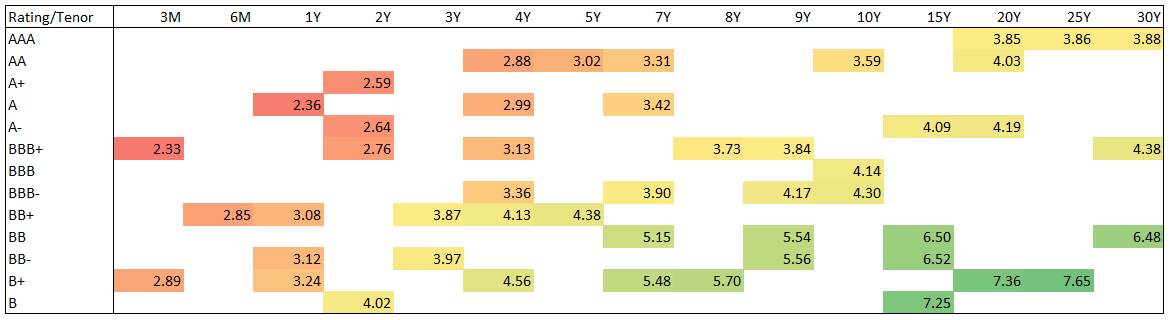}\\
  \caption{A sample full/sparse yield matrix (Source: Bloomberg).}\label{full_sparse_mat}
  \end{center}
\end{figure}

In many fixed income markets, market liquidity is insufficient to facilitate price discovery and as a result, it can be difficult to find bond yields for all rating and tenor pairs. Therefore, on each day, the quotes for the bond yields for various ratings and tenors can be considered as a sparse matrix (see Figure \ref{full_sparse_mat}). 

The central idea of the present paper is to design an unsupervised machine that learns the salient features of corporate bonds yield curves by observing a sufficient number of historical examples in a liquid market, and then uses the learned shapes to fill in the missing yields in the illiquid market. Roughly speaking, the machine is a tool for interpolation/extrapolation, however, it also incorporates its memory of typical yield curve shapes.

Our main insight in approaching the problem is to consider the sparse yield matrix as a corrupted image. Therefore the goal is to reconstruct the full yield matrix in the same way images are reconstructed from noisy ones. 

For this purpose, several algorithms that have been successfully employed for various tasks in computer vision were considered, all of which belong to the class of \textit{inpainting algorithms} \cite{TV1,TV2,MonotoneInterp}. The first algorithm considered is the denoising autoencoder (DAE) which was first proposed for the purpose of better training deep neural networks on images in the pre- RELU \footnote{RELU stands for Rectified Linear Unit which is the most common activation function of hidden layers in modern deep learning architectures.} era of deep learning (see \cite{DeepBook} and \cite{Bengio} for background). Several types of architectures of DAEs were tested, in order to achieve highest performance given the limited size of the data (see Sections \ref{AE_arch} and \ref{experiments}).

Also considered in this paper are the \textit{total variation inpainting} (TV) and the \textit{thin plate spline} (TPS) algorithms. One important difference between these algorithms and DAE is that the former algorithms are calibrated in a point-in-time format. More precisely, DAE tries to learn the empirical features of the yield curves by training over historical observations, whereas both TV and TPS try to fill in the given sparse matrix using only the structure present in the matrix and without any knowledge of history. 

It was observed that while the performance of three algorithms were about on par for uniform sparse yield matrices (see Section \ref{experiments}), for the cases of block sparsity, the DAE exhibits superior generalization properties.

The paper is organized as follows. In Section \ref{problem}, the setup of the problem is explained. In Section \ref{theory} , a high level description of the DAE algorithm and various network architectures for DAEs is given. For the sake of completeness, the details of the TV and TPS algorithms are included. In Section \ref{experiments}, the details of the used data and algorithms implementation is presented. A summary of the experimental results is provided in Section \ref{results}.

%The title of your section 2
\section{Problem Definition}\label{problem}

The problem setup is as follows. Consider a fixed income trader who must provide quotes on prices for a variety of illiquid corporate/sovereign bonds for different ratings and tenors in a specific market. The trader only has access to bond yields for a few anchor points on the rating/tenor grid and she is tasked to complete the rating/tenor matrix based on the given points. 

\begin{figure}[htp]
\begin{center}
  % replace aims_logo.eps by your figure file name
  \includegraphics[width=4in]{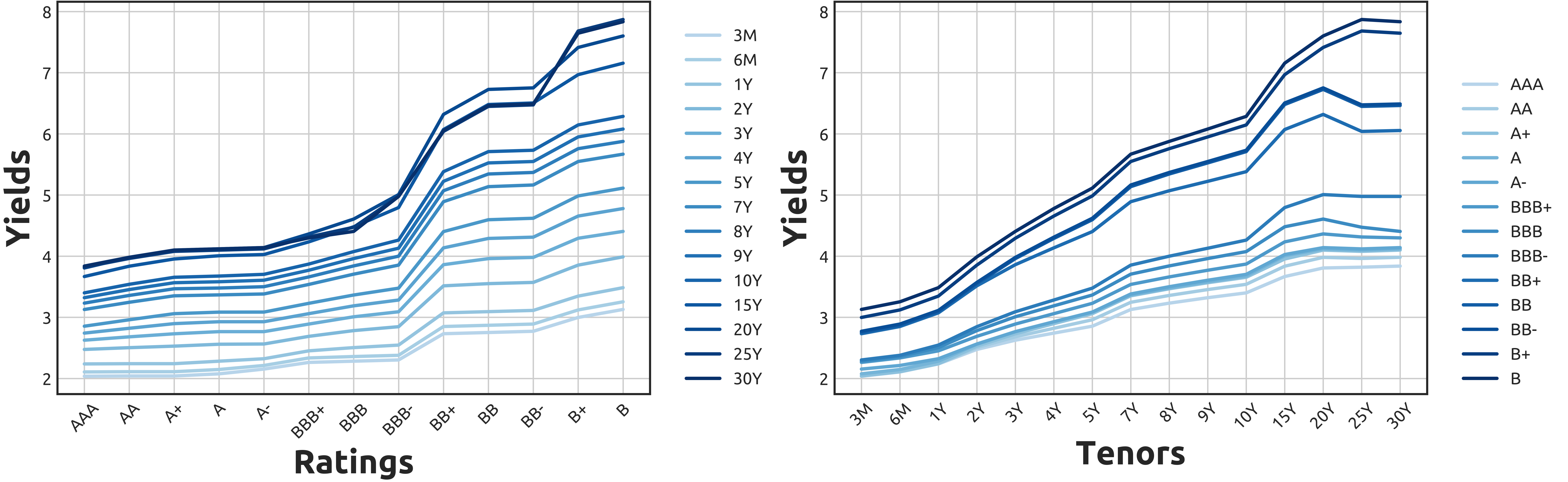}\\
  \caption{A typical yield curve structure (source:~Bloomberg)}\label{yields}
  \end{center}
\end{figure}

This is a problem of interpolation/extrapolation under market no-arbitrage constraints. These constraints are embodied collectively as the features present in the shape of the curves. For instance, in a normal market environment, the yield curves for each rating are upward sloping, with longer term interest rates being higher than shorter term (see Figure \ref{yields} for an example). The curve might exhibit other features such as humps and mixes etc., based on the supply and demand for the instruments. In typical situations for our problem, some of the yields at the short end of the curves for only some of the ratings are known. 

There exist numerous approaches for interpolation of the yield curve along the tenor coordinate (see \cite{WilmPaper} for a survey of the methods). However, the task we are interested in is to fill in a sparse matrix grid along both the rating and the tenor coordinates, therefore a two dimensional generalization of the above described methods is needed.

Among the point-in-time two dimensional interpolation methods, TV (\cite{TV1}, \cite{TV2}) and TPS algorithms from the image processing domain can be applied (see Sections \ref{TV} and \ref{TPS} for details).
The TPS surfaces, \cite{ESLBook},\cite{TPS1}, \cite{TPS2} are natural generalization of the 1-D smoothing splines to 2-D, and are effective tools for filling in a 2-D grid and transforming it to a surface while the oscillatory behaviour of the interpolating surface can be controlled (see Section \ref{TPS} below). 

As noted in \cite{WilmPaper}, spline interpolators, and more generally, the point-in-time interpolation methods, do not necessarily produce the empirically expected features in the yield curve. This is also true for the TV and TPS algorithms. Therefore, the raw outputs of the interpolation algorithm must undergo an additional modification step in order to produce the desired curve features (such as monotonicity, concavity, etc.). Our proposed unsupervised method will be able to infer the standard features (and perhaps other properties) directly from the training data it has observed.

\section{Theoretical Background}\label{theory}

In this section we provide a brief presentation of the theoretical foundations for the algorithms implemented in this paper. We define a class of unsupervised neural networks called Autoencoders (AEs). We specify the sub class of this family of algorithms called Denoising Autoencoders (DAE). We also describe both the Fully Connected Neural Network (FCNN) and Convolutional Neural Network (CNN) architectures for these algorithms. We also introduce the TV and TPS algorithms which are the challenger models to DAE in this paper. We refer the reader to \cite{DeepBook} for details of NNs, CNNs and autoencoders. The reader can consult \cite{ESLBook}, \cite{TPS2}, \cite{TV1} and \cite{TV2} for more details on TV and TPS algorithms.

\subsection{Total Variation Inpainting}\label{TV}
The contribution of this paper is to formulate the yield matrix interpolation/extrapolation as an image inpainting (a.k.a image completion) problem, for which the goal is to recover the missing parts of the image. Inpainting problems are generally formulated as an optimization that minimizes the summation of the recovery error and a regularization term which models the global behaviour of the matrix, 
\begin{equation}
\argmin_f \sum_i \| f(X_i) - y_i \|_2^2 + \lambda R(f)  
\end{equation}

where $y$ is the observed matrix, $f(X)$ is the recovered matrix, $\lambda>0$ is the regularization weight, and $R(f)$ is the regularization term. Considering the local smoothness of the images, a very popular class of the inpaintings uses the first order gradient of the images in vertical and horizontal directions, i.e. Total Variation (TV) inpainting:
\begin{equation}
\argmin_f \sum_i \| f(X_i) - y_i \|_2^2 + \lambda \sum_{i=0}^{N-1}\sum_{i=0}^{N-1} |\nabla f(X)_{i,j}|.  
\end{equation}

\subsection{Thin Plate Spline}\label{TPS}
A well known problem in the TV based algorithms is the tendency to remove the high frequency changes in image features. Therefore, other complementary algorithms are proposed to improve the quality of the recovered images. The class of dictionary learning based techniques can learn the high frequency structures in the same class of the images. The learned dictionaries will then be used in sparse coding to recover the missing high frequency contents of the image (see \cite{DictLearn} and references therein).  

However, by looking at the structure of the Rating-Tenor surface (Figure \ref{yields}), it can be seen that the yield matrix has a smooth semi-quadratic structure. This encourages the application of higher order regularization. A possible choice, found to be useful in this case, is the sum squared of Hessian matrix (i.e. second order differentiation), a method known as Thin Plate Spline (TPS).

The TPS surfaces are 2-D generalizations of 1-D cubic splines (see \cite{ESLBook}, \cite{TPS1}, \cite{TPS2}). TPS is an interpolation algorithm for the grid points $\{(X^i, y^i)\}_{i=1}^{m}$ where $X^i=(x^i_1, x^i_2)\in\mathbb{R}^2$. The spline surface $f(x_1,x_2)$ is constructed such that it passes the grid points as closely as possible while exhibiting a controlled amount of oscillation. This Bias/Variance optimal surface is obtained by minimizing the following action functional over an appropriate function space (see \cite{TPS2}).
\begin{equation}\label{action}
S[f] = \sum_{i}\|f(X_i)-y_i\|^2 +\lambda\int|H(f)|^2dx.
\end{equation}
Here, $|H(f)|^2$ denotes the sum of square entries of the Hessian matrix of $f$ and $\lambda\in\mathbb{R}^+$ is a regularization parameter. The first term is a measure of fitting error and the integral in the second term measures the oscillating behaviour of the interpolation surface. 

The TPS solution is the minimizer of the above defined action. Applying the Euler-Lagrange recipe for finding the minimum of the functional in (\ref{action}) one obtains a 4th order PDE with certain orthogonality conditions on the space of solutions (see \cite{ESLBook}, \cite{TPS1}). The solution in the 3D euclidean space (2D TPS) is given by the following closed form expression,
\begin{equation}
f(x)=\sum_{i=1}^m a_i G(X,X_i)+(b_0+b_1x_1+b_2x_2).
\end{equation}
In practice, in the 3D case one can set $G(X,s)=u(|X-s|)$, where $u(x)=x^2\log x$. Using the constraints $f(X_i) =y_i$, one obtains the following linear system,
\begin{equation}\label{eqsystem}
Y= Ma+Nb,
\end{equation}
where $M$ is an $m\times m$ matrix with the entries $M_{ij}=G(X_i,X_j)$ and $N$ is an $m\times 3$ matrix with the rows $[1~~X_i^T]$. The system is subject to the orthogonality condition (see \cite{TPS1}) $N^Ta=0$. It turns out that the matrix $M$ is non-singular and the system (\ref{eqsystem}) can be solved by,
\begin{equation}
b=(N^TM^{-1}N)^{-1}N^TM^{-1}Y,
\end{equation}
\begin{equation}
a=M^{-1}(Y-Nb).
\end{equation}

The regularization weight $\lambda$ must be set in advance. In practice, $\lambda$ is chosen through a hyper-parameter tuning process such as $k-$fold cross validation (see \cite{ESLBook} and \cite{fields}).

\subsection{Autoencoders}
The references \cite{DeepBook} and \cite{Bengio} provide the basics of neural networks, we follow their concepts and notation closely.
Autoencoders can be described as a class of unsupervised learning algorithms which are designed to transfer the inputs to the output through some (non-linear) reproducing procedure. They are called unsupervised since the algorithm only uses the inputs $X$ for learning. An autoencoder transforms an input vector $x\in\mathbb{R}^d$ to a hidden representation $h(x)\in\mathbb{R}^{d'}$ via a mapping $h(x) = g(W x + b)$ (see Figure \ref{AE} below). $W$ is a $d'\times d$ weight matrix and $b$ is called a bias vector. The mapping $g(.)$ is also called the \textit{encoder} part of the autoencoder. 

The \textit{decoder} part of the algorithm then maps the obtained hidden representation back to a “reconstructed” vector $z\in\mathbb{R}^d$ via a mapping $z = k(W^*h(x)+c)$. 

\begin{figure}[htp]
\begin{center}
  % replace aims_logo.eps by your figure file name
  \includegraphics[width=4in]{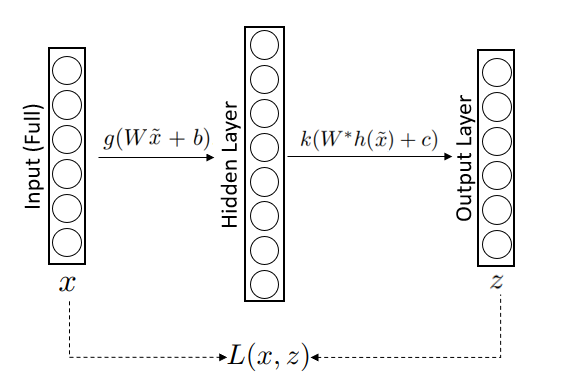}\\
  \caption{ A general autoencoder architecture.}\label{AE}
  \end{center}
\end{figure}

Therefore, observation $x_i$ is first mapped to a corresponding $h(x_i)$ and then to a reconstruction $z_i$. If we define $\theta=\{W,b\}$ and $\theta'=\{W^*,c\}$, the optimal weights are found by solving,
\begin{equation}
    \argmin_{\{\theta,\theta'\}}\frac{1}{N}\sum_{i=1}^{N}L(x_i,z_i).
\end{equation}
Here, $L(x,z)$ is a chosen \textit{loss function}. A principal recipe in statistical learning for density estimation \cite{Vapnik} can guide us in choosing the cost function for training the autoencoder. In general, one starts with choice of a joint distribution of the data $p(x|\theta)$, where $\theta$ is the vector of distribution parameters. Next, one defines the relationship between $\theta$ and $h(x)$ and sets up a loss function $L(k\circ g(x))=-\log p(x,\theta)$, where $k\circ g(x)=z$ is the autoencoder functional form. For instance, the choice of square error loss $L(x,z)=\|x-z\|^2$ is equivalent to the choice of a Gaussian distribution with mean $\theta$ and the identity covariance matrix for the data,
\begin{equation}
    p(x,\theta) = \frac{1}{2\pi^{N/2}}exp(-\frac{1}{2} \sum_{i=1}^{N}(x_i-\theta_i)^2),
\end{equation}
where $\theta=c+W^*h(x)$ (see \cite{DeepBook}).

\subsection{Denoising Autoencoders}
The basic idea of an autoencoder as explained above is to minimize an expected loss function $\mathbb{E}[L(x,k\circ g(x)]$. Here, the loss function penalizes the \textit{dissimilarity} of the function $k\circ g(x)$ from $x$. This may drive the functional form to be merely an identity mapping $k\circ g(x)=x$ if the algorithm architecture allows for it. In other words if, the hidden layer $h(x)$ of the autoencoder is wider than the input vector (overcomplete), the algorithm would just copy the inputs to the hidden layer and hence learn no meaningful features of the inputs.

To avoid this situation, a denoising autoencoder (DAE) can be trained. (This section follows the exposition in \cite{Bengio} closely; see \cite{DeepBook} and references therein for more details.) The idea is to minimize the expected loss $\mathbb{E}[L(x,k\circ g(\tilde{x}))]$ where $\tilde{x}$ is a \textit{noisy} version of $x$. The DAE must therefore learn sufficient features in order to salvage the original data from the corrupted version rather than simply copying their input.

The process starts by first injecting noise (masking) to the initial input $x$ to obtain a partially corrupted copy $\tilde{x}$ via a stochastic procedure (so, $\tilde{x}\sim q_D(\tilde{x}|x$). In \cite{Bengio} the following corruption process is implemented: for each input $x$, a fixed proportion $\nu$ (fixed in advance) of the coordinates are randomly set equal to $0$ while the others are left untouched.\footnote{An alternative way is to use Gaussian additive noise whose variance is set via  hyper-parameter tuning (see \cite{DeepBook}).} 

\begin{figure}[htp]
\begin{center}
  % replace aims_logo.eps by your figure file name
  \includegraphics[width=4in]{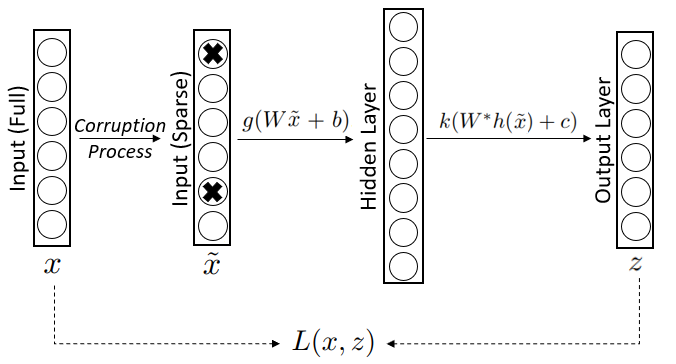}\\
  \caption{ A DAE corrupts and then reconstructs the input.}\label{DAE}
  \end{center}
\end{figure}

The corrupted copy of the input $\tilde{x}$ is then fed into a regular autoencoder (see Figure \ref{DAE}). It is important to notice that the parameters of the model are trained to minimize the average reconstruction error $\mathbb{E}[L(x, z)]$ not $\mathbb{E}[L(\tilde{x}, z)]$, namely to obtain $z$ as close as possible to the
uncorrupted input $x$. Here, the output $z$ is related deterministically to $\tilde{x}$ rather than $x$.

\subsection{Architectures for Autoencoders}\label{AE_arch}

 The encoder and decoder parts of an autoencoder are feedforward neural networks with variable architectures (see \cite{Bengio} for definitions and details).
 The Figure \ref{DAE} above demonstrates the fully connected neural network (FCNN) for the DAE in this paper (see Section \ref{training}). It is seen in the figure that the input filtered through the encoder to produce the code. The decoder then reconstructs the output based on the code.
 
  The decoder architecture depicted above in Figure \ref{DAE} is the mirror image of the encoder, this is not a requirement but a popular choice of architecture. The only requirement is the dimensions of the input space and the output space must be identical. 
 
\begin{figure}[htp]
\begin{center}
  % replace aims_logo.eps by your figure file name
  \includegraphics[width=4in]{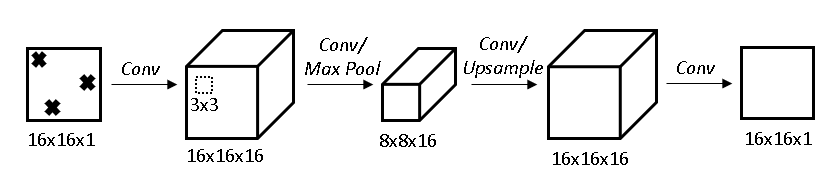}\\
  \caption{ The 1-Channel CNN network trained on historical rate surfaces.}\label{CNN}
  \end{center}
\end{figure}

As emphasized in \cite{conv_paper}, FCNN architectures for autoencoders and DAEs both ignore the 2-D structure of an image (2-D structure of the rating/tenor surface in our case). Hence, the most successful models in image pattern recognition capture the localized features that repeat themselves in the input (see \cite{conv_paper} and references therein). These algorithms employ the CNN architectures for both the decoder and encoder parts of autoencoders (see Figure \ref{CNN} and \cite{DeepBook} for details on CNNs).

Two types of CNN architectures were trained for DAE. The first CNN- DAE network has only a 1-channel input, namely the masked rate surface (Figure \ref{CNN} above). The second architecture has a 3-channel input which is a tensor of size $16\times16\times3$. The first layer of the stack is the masked rate surface and the other two stacks encode the coordinates of the the rates on the grid. 

This architecture was inspired by the position embedding techniques employed in other applications of deep learning such as sequence to sequence learning (see \cite{PECNN}). The basic idea is that position embedding channels give the network a sense of which area of the sparse surface in the input it is dealing with. It was observed that the CNN- DAE with position embedding is better at capturing the monotonicity of rates in rating (see Section \ref{perf}).

\section{Experimental Results} \label{experiments}
\subsection{Data Description}

The data are collected from Bloomberg (under the Western University's academic licence) and consist of mid-yields of corporate industrial bond indices. There are 13 indices (AAA, AA, A+, A, A-, BBB+, BBB, BBB-, BB+, BB, BB-, B+, B), and each index provides yields at 15 tenors (3m, 6m, 1y, 2y, 3y, 4y, 5y, 7y, 8y, 9y, 10y, 15y, 20y, 25y, 30y), resulting in a 2D surface with 195 points for each observation date. The yields are computed by Bloomberg from constituent bonds and are taken as given. The data are daily and range from Jan 29, 2018 to April 27, 2018 (63 observations).

Yields at 20, 25 and 30 years are unavailable for the BB+, BB, BB-, B+ and B indices. These missing points are populated by finding the spread between the missing tenor and the 15 year tenor for generic double and single B industrial corporate indices, and assuming that the same spread applies to the corresponding more granular rated indices.

It is observed that the yields are weakly monotonic in rating (with the exception of B+ and B indices at the long end of the curve, for a single observation). Yields are also monotonic in tenor over the period observed, except in the very long end of the curve for some indices. 

\subsection{Fitting the Total Variation}
To recover the missing values with TV, the \texttt{TVL2Denoise} function from SParse Optimization Research COde (SPORCO) \cite{SPORCO} package was used. SPORCO is a Python package for solving optimization problems with sparsity-inducing regularization. These consist primarily of sparse coding and dictionary learning problems, including convolutional sparse coding and dictionary learning, but there is also support for other problems such as Total Variation regularization and Robust PCA. It uses Alternating Direction Method of Multipliers (ADMM) \cite{ADMM} to solve the TV optimization. The TV-inpainting algorithm was fitted to each of the 70 sparse matrices in the test set and then full matrix was predicted and compared with actual full test matrices to derive the relevant error metrics. 

\subsection{Fitting the Thin Plate Spline}
The \texttt{Tps} functionality in the $R$-package \texttt{fields v9.6} was used in order to fit the TPS surfaces. We refer the reader to \cite{fields} for the details and set up of the functionality. The TPS algorithm was fitted to each of the 70 sparse matrices in the test set and then full matrix was predicted using the calibrated tps parameters. The results were then compared with actual full test matrices to derive the relevant error metrics. 

\subsection{Training the Autoencoder}\label{training}

Training and test data for the neural networks are constructed as follows. First, the yields are scaled such that the maximum yield is 1. This aligns the model inputs and outputs with the range of the sigmoid activation function. Then, 10 percent of the observations are held out for testing. Since the purpose of the algorithm is to reconstruct a rate surface from known inputs, rather than predict a rate surface out of time, the test set is selected at random from the data.

Finally, to provide the algorithm with additional information for training, each observation from both the training and test sets is repeated 10 times and subjected to a randomized corruption procedure. Two types of corruption (masking) was used for the purpose of testing. In the first method, each training matrix is masked uniformly. More precisely, following (\cite{Bengio}), a fixed number of elements from each example are chosen at random and their value is forced to 0, creating sparse rate surfaces. 

In the second method we also masked three quadrant of the uniformly sparse matrix and only left the top left quadrant, untouched. This was motivated by the use cases in practice where in illiquid markets, usually liquid trades exist only for highly rated and short maturity fixed income instruments. Each neural network is then trained to reconstruct complete rate surfaces using the sparse surfaces as inputs.

The neural networks are implemented in Keras with TensorFlow backend. Adam optimization with standard parameter values and mean square error loss is used for both networks. The Python package \texttt{Hyperas}, with built-in Tree-of-Parzen-Estimators (TPE) algorithm (\cite{Hyperopt}), is used to conduct hyperparameter optimization for the learning rate, decay rate, batch size, number of nodes in the FCNN hidden layer, and number of CNN layers and filters. The Rectified Linear Unit (ReLU) is used as the activation function for hidden layers, and the sigmoid as the activation function for the output layer. Batch normalization is used after each hidden layer activation in both networks.

The FCNN uses a single overcomplete hidden layer (see Figure.\ref{DAE}). Implicit in the structure is the idea that the network must learn relationships between each pair of ratings in the surface, no matter the distance between them in tenor/rating space. However, economic intuition suggests that adjacent yields on the surface will be strongly correlated. The CNN captures this feature in its network architecture, which consists of several layers of convolutions with 3x3 filters and pooling, followed by convolutions and upsampling to restore the rate surface dimensions. As such, the CNN has considerably fewer parameters than the FCNN.

\subsection{Monotonicity and Smoothness of the Autoencoder}
One of the important features of the yield curves is the monotonicity along the ratings, meaning the yield of a higher rated bond is always smaller than the yield of a lower rate bond. In addition, the yield curves change smoothly along both tenor and ratings (see Figure \ref{yields}). It was observed that, using the rate-tenor surface as the training inputs to the DAE, it won't learn the monotonicity and smoothness very well. 

We hypothesize that this is due to the application of multi-layer CNN/FCNN architecture, which reduces the amount of information about the ratings and tenors provided to deeper layers of the network. To test this hypothesis, two extra pieces of input information are added as extra channels to the input rating-tenor surface. One channel contains increasing values along the ratings while the values are kept constant along tenor. The second channel increases along tenor and is kept constant along ratings. These two extra channels provide information about the tenor and ratings to the DAE. 

As was mentioned in \ref{AE_arch}, in other areas of deep learning applications, such as sequence to sequence learning, similar technique is used in order to help the network get a sense of the position or coordinate in the input tensor (see \cite{PECNN}). In the following, we denote the CNN architecture with position embedding channel by CNN/PE. The DAE trained with ratings-tenor surface only (one channel) will be referred to as CNN. 

\subsection{Performance Testing}\label{perf}

\begin{figure}[htp]
	\begin{center}
		% replace aims_logo.eps by your figure file name
		\includegraphics[width=4in]{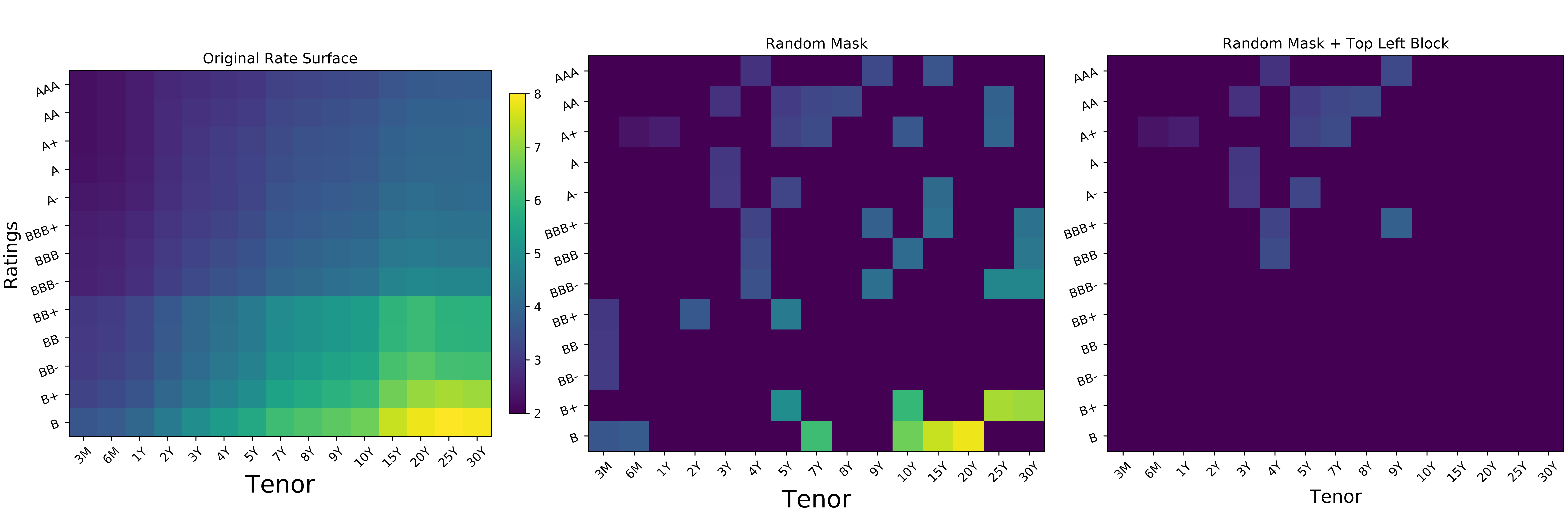}\\
		\caption{An example of a rate surface (left), uniformly masked (middle), and block sparse masked rate (right) surfaces.}\label{Masks}
	\end{center}
\end{figure}

\begin{figure}[htp]
\begin{center}
  % replace aims_logo.eps by your figure file name
  \includegraphics[width=5in]{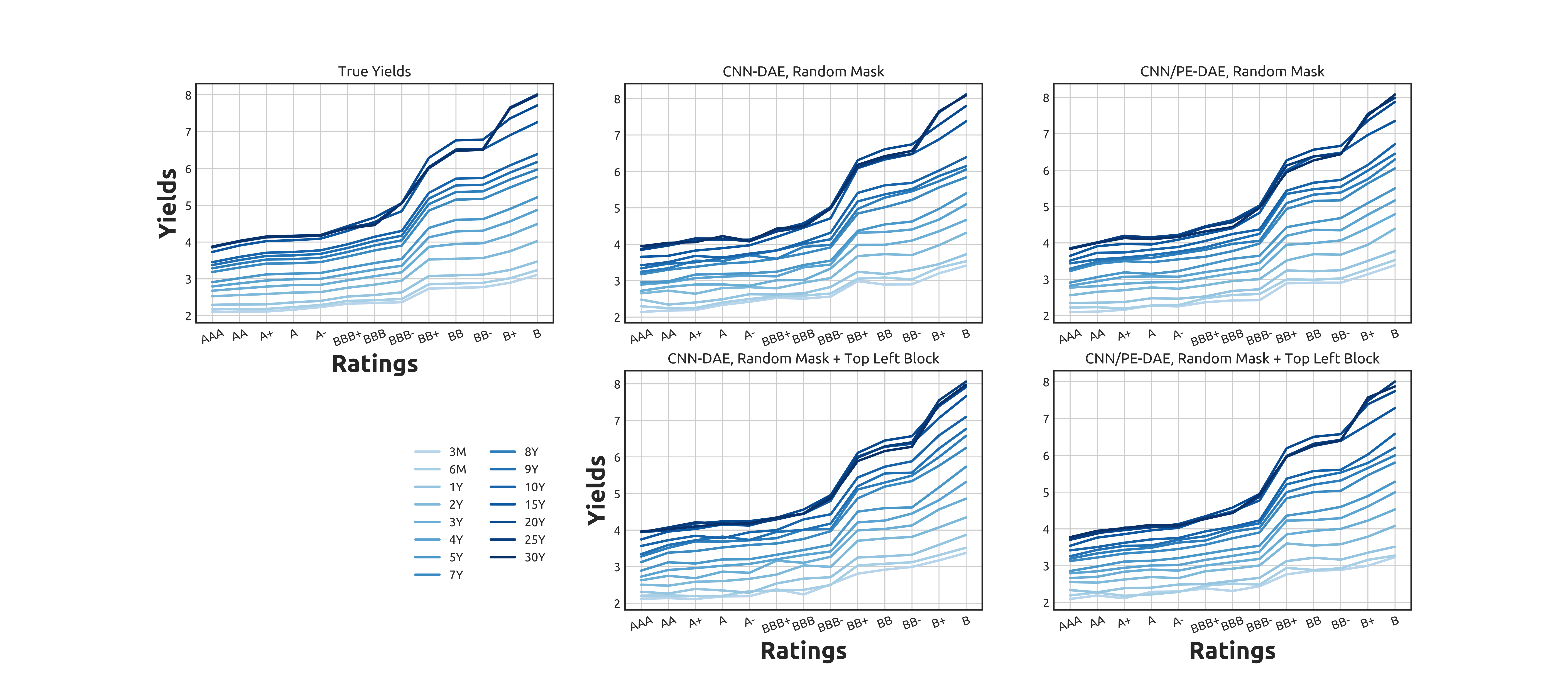}\\
  \caption{ Reconstructed yield curves for a uniformly masked (top) and block masked (bottom) test example by DAE with CNN architectures.}\label{CNN_mon}
  \end{center}
\end{figure}

\begin{figure}[htp]
\begin{center}
  % replace aims_logo.eps by your figure file name
  \includegraphics[width=4in]{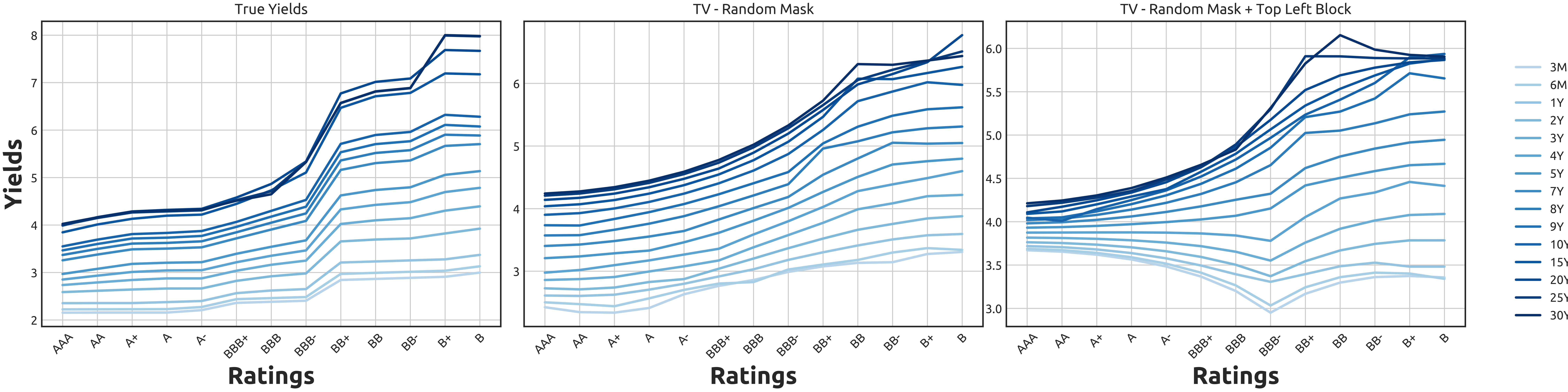}\\
  \caption{ Reconstructed yield curves for a uniformly masked test example (middle) and block masked (right) by TV algorithm.}\label{TV_mon}
  \end{center}
\end{figure}

Our test set consists of 70 examples. There are two types of test measure, one for the uniform masking and one for the block masking (top left quadrant, see Figure.\ref{Masks}). For the uniform masking, $25\%$ sparsity is produced. Table \ref{accuracy_unif} and \ref{accuracy_block} contain the accuracy of the considered algorithms on the corresponding test set. Figures \ref{CNN_mon} and \ref{TV_mon} visualize the outputs of DAE and TV algorithms on a chosen test example.

The Mean Absolute Error (MAE) and Root Mean Square Error (RMSE) are calculated over all points on the rate surface for all test set examples, and the values in percent are computed relative to the true observed yields.Note that CNN refers to the CNN-DAE with 1 channel input and CNN/PE refers to the CNN-DAE with position embbeding.

\begin{table}[!hbt]
\center{
\begin{tabular}{|c|c|c|c|c|c|}\hline
Metric & TV & TPS & FCNN & CNN &CNN/PE\\ \hline\hline
MAE (bps) & 21.43 & 11 & 8 & 10 & 8.85\\
MAE (percent)& 6.51 & 3 & 2 & 3 &2.29\\
RMSE (bps) & 31.86 & 18 & 11 & 13 & 14.66\\
RMSE (percent) & 8.6 & 4 & 3 & 3 & 4.14\\
\hline
\end{tabular}
}
\vskip 0.25cm
\caption{Test set performance, uniform masking with $\nu$ = 0.75.}\label{accuracy_unif}
\end{table}

\begin{table}[!hbt]
\center{
\begin{tabular}{|c|c|c|c|c|c|}\hline
Metric & TV & TPS & FCNN & CNN &CNN/PE\\ \hline\hline
MAE (bps) & 31.12 &20.07& 10.01& 13.61 & 10.96\\
MAE (percent)& 18.16& 12.41 & 2.7 & 3.59 & 2.97\\
RMSE (bps) & 40.21 & 25.31& 13.69 & 17.11 & 14.66\\
RMSE (percent) & 26.47 & 14.08 & 3.78 & 4.60 & 4.14\\
\hline
\end{tabular}
}
\vskip 0.25cm
\caption{Test set performance, block masking.}\label{accuracy_block}
\end{table}

It is seen that the overall performance of the algorithms is better on uniformly masked data, which is reasonable given that uniform sparsity still provides some means for the algorithm to guess the interpolating pieces. The TV algorithm shows the worst performance. Note that, while the performance of TPS, FCNN, CNN and CNN/PE are more or less on par on the uniformly sparse examples, the DAE architectures perform noticeably better on the block masked sparse data.

The FCNN and CNN procedures were also run on a proprietary data set with a substantially longer history, including a variety of different surface shapes. For this data set, the CNN outperforms the FCNN, suggesting that it may have superior flexibility in learning a variety of tasks.

The algorithms were also examined to see how much of the monotonicity along the ratings feature in the training set is preserved. The metric used was a simple difference method along the rating directions to identify the negative slopes. The following tables summarizes the monotonicity preserving performance of the algorithms.

\begin{table}[!hbt]
\center{
\begin{tabular}{|c|c|c|c|c|}\hline
TV & TPS & FCNN & CNN &CNN/PE\\ \hline\hline
3.60 & 3.11 & 6.4 &10.25 &5.43\\
\hline
\end{tabular}
}
\vskip 0.25cm
\caption{Percentage of monotonicity violations in the test set, uniform masking.}\label{accuracy_block}
\end{table}

%%%%%%%%%%%%%%%%%%%%%%%%%%%%%%%%%%%%
\begin{table}[!hbt]
\center{
\begin{tabular}{|c|c|c|c|c|}\hline
TV & TPS & FCNN & CNN &CNN/PE\\ \hline\hline
23.56 & 18.91 &9.6 &12.77 &7.50\\
\hline
\end{tabular}
}
\vskip 0.25cm
\caption{Percentage of monotonicity violations in the test set, block masking.}\label{accuracy_block}
\end{table}

\section{Summary of Results}\label{results}

This paper has demonstrated a novel financial application of well known algorithms in the image recognition literature. FCNN and CNN DAEs are shown to be capable of extracting features from liquid markets. Assuming that these same features are present in illiquid markets, the algorithms can be used to estimate missing information. This paper provides an example for corporate bonds, but similar approaches are likely to be fruitful in other areas such as equity volatility surfaces.

%For acknowledgements section, please don't number the section, please begin it with \section*{Acknowledgements}
%\section*{Acknowledgments} We would like to thank you for \textbf{following
%the instructions above} very closely in advance. It will definitely
%save us lot of time and expedite the process of your paper's
%publication.

% You may incorporate your references as follows in your main tex file.
% Using BibTex is not recommended but can be handled.

\medskip
% The data information below will be filled by AIMS editorial staff
Received xxxx 20xx; revised xxxx 20xx.
\medskip

\end{document}